\documentclass{aa}  

\PassOptionsToPackage{usenames,dvipsnames}{xcolor}
\usepackage[varg]{txfonts}
\usepackage{graphicx}
\usepackage[colorlinks=true,citecolor=blue]{hyperref}
\usepackage{soul}

\defcitealias{2018A&A...620A..78S}{SK18}
\defcitealias{2019A&A...631A..96I}{IL19}

\newcommand{\rmd}{\mathrm{d}}

\newcommand{\uncite}[1]{(#1)}
\soulregister{\uncite}{1}
\newcommand{\unref}[1]{(#1)}
\soulregister{\unref}{1}
\newcommand{\unlabel}[1]{(#1)}
\soulregister{\unlabel}{1}

\begin{document} 

\title{Cluster counts III.  $\Lambda$CDM extensions and the cluster tension}
\titlerunning{$\Lambda$CDM extensions and the cluster tension}
\author{Ziad Sakr\inst{1,2}
\and Stéphane Ili\'c\inst{3,1}
\and Alain Blanchard\inst{1}}
\authorrunning{Sakr et al.}
\institute{IRAP, Université de Toulouse, CNRS, CNES, UPS, Toulouse, France
\and
Universit\'e St Joseph; Faculty of Sciences, Beirut, Lebanon
\and
 Universit\'e PSL, Observatoire de Paris, Sorbonne Universit\'e, CNRS, LERMA, F-75014, Paris, France \\
\email{\scriptsize zsakr@irap.omp.eu, stephane.ilic@gmail.com, alain.blanchard@irap.omp.eu}
}
   
   \date{Received ; accepted }

  \abstract{
Despite the success of the Lambda cold dark matter ($\Lambda$CDM) cosmological model, current estimations of the amplitude of matter fluctuations ($\sigma_8$) show an appreciable difference between its value inferred from the cosmic microwave background (CMB) angular power spectrum ($C_{\ell}$) and those obtained from cluster counts. Neutrinos or a modification of the growth of structures had been previously investigated as the possible origin of this discrepancy. In this work we examine whether further extensions to the $\Lambda$CDM model could alleviate the tension. To this end, we derived constraints on the parameters subject to the discrepancy, using CMB $C_{\ell}$ combined with cluster counts from the  Sunyaev--Zel'dovich (SZ) sample with a free dark energy equation of state parameter, while allowing the cluster mass calibration parameter $(1-b)$ to vary. This latter is degenerate with $\sigma_8$, which translates the discrepancy within the  $\Lambda$CDM framework into one between $(1-b) \sim0.6$, corresponding to constraints on $\sigma_8$ obtained from CMB, and $(1-b)\sim0.8$, the value adopted for the SZ sample calibration. We find that a constant $w$, when left free to vary along with large priors on the matter density ([0.1,1.0]) and the Hubble parameters ([30,200]), can reduce the discrepancy to less than 2$\sigma$ for values far below its fiducial $w$ = -1. However, such low values of $w$ are not allowed when we add  other probes like the  baryonic acoustic oscillation (BAO) feature angular diameter distance measured in galaxy clustering surveys. We also found, when we allow to vary in addition to $w$ a modification of the growth rate through the growth index $\gamma$, that the tension is alleviated, with the $(1-b)$ likelihood now centred around the Planck calibration value of $\sim$ 0.8. However, here again, combining CMB and cluster counts with geometrical distance probes restores the discrepancy, with the  $(1-b)$ preferred value reverting back to the $\Lambda$CDM value of $\sim$ 0.6. The same situation is observed when introducing, along with $w$ and $\gamma$, further extensions to $\Lambda$CDM (e.g. massive neutrinos), although these extensions  reduce the tension to 2$\sigma$, even when combined with BAO datasets. We also explore other common extensions by comparing two cases: allowing a dynamical $w$ following a CPL parametrisation in addition to a constant growth index, and when the growth index is expanded through a second parameter $\gamma_1$ along with a constant $w$. In the former we reach the same conclusions as with the case of a constant $w$ and $\gamma$, where the discrepancy was alleviated only if we do not constrain $w$ by BAO, while in the latter case, we observe that introducing $\gamma_1$ drives $(1-b)$ towards lower values that would instead increase the discrepancy on $\sigma_8$. We conclude that none of these common extensions to $\Lambda$CDM is able to fix the discrepancy and a misdetermination of the calibration factor is the most preferred explanation. Finally, we investigate the effect on our posteriors from limiting the Hubble constant priors to the usual common adopted range of [30,100].}


\maketitle


\section{Introduction}\label{sec:IntrO}

Cluster counts are considered  a powerful tool for constraining cosmological parameters \citep{1992A&A...262L..21O,2005ASPC..339..140M,2009ApJ...692.1060V,2011ARA&A..49..409A}. Their abundance is primarily sensitive to two parameters, the matter density $\Omega_{\rm m}$ and the current amplitude of matter fluctuations, $\sigma_{\rm 8}$  \citep{1997ApJ...485L..53B,1998A&A...332L..49B}. The value of the latter, constrained from the Planck mission \citep{2016A&A...594A..13P} assuming the  Lambda cold dark matter ($\Lambda$CDM) model, was found to be  in tension with the value determined from Sunyaev--Zel'dovich (SZ) cluster counts at lower redshifts \citep{2014A&A...571A..29P}. Similar discrepancies, although with different amplitudes, were also found when using other large-scale structure local probes, such as weak gravitational lensing detected clusters from the Canadian Cluster Comparison Project (CCCP) \citep{2015MNRAS.449..685H}, Sunyaev Zel'dovich identified galaxy clusters from the South Pole Telescope (SPT) survey  \citep{2016ApJ...832...95D}, and  gravitational lensing cluster mass measurements from the Cluster Lensing And Supernova and Hubble (CLASH) survey  \citep{2017A&A...604A..89P}, while only few, such as the  Weighing the Giants (WtG) weak lensing cluster survey \citep{2014MNRAS.439...48A}, claim a more appropriate selection function and  yield values of $\sigma_8$ in agreement with those from the Planck mission. Even though the early release from the large-scale Dark Energy Survey (DES) \citep{PhysRevD.98.043526} or more recent ones \citep{DES:2021bvc} with constraints from cosmic shear measurements and the latest Planck cosmic
microwave background (CMB) observations \citep{2020A&A...641A...6P} have shown a lesser tension,  the discrepancy is still present, especially when the   CMB $\sigma_8$ derived constraints are compared to those from the latest galaxy clusters surveys \citep{2019arXiv191013548K,2019ApJ...878...55B,2020A&A...634A.127A,2020ApJ...901...90A}.\\
The origin of this tension is still under investigation. On the CMB side, the high accuracy achieved by the Planck mission in measuring the CMB temperature angular power spectrum in the main release \citep{2016A&A...594A..13P}, or the final polarisation spectra in \cite{2020A&A...641A...6P}, as well as other CMB experiments like the latest release of the Atacama Cosmology Telescope (ACT) combined with the Wilkinson Microwave Anisotropy Probe (WMAP) \cite{ACT:2020gnv}, leaves small room for any large systematics that could be responsible for a misdetermination of the cosmological parameters by this mission. Moreover, an analysis by \citet{Motloch:2019gux} of the final release of the Planck satellite data on the anomaly observed between the lensing gravitational potential and the lensing obtained from the smoothing of the acoustic peaks in the temperature and polarisation power spectra found that marginalising over this component cannot substantially resolve the $\sigma_8$ tension with low-redshift measurements.\\
This is not the case for the cluster counts, where substantial systematics remain when trying to relate the observable abundance of clusters to the theoretical mass function. This is suggested from the scatter found between the values of the calibrations of the different scaling relations connecting cluster mass to observable quantities, such as X-ray emission from cluster hot gases, cluster richness, other optical properties, or the SZ effect. This uncertainty in the determination of the mass calibration parameter could have a strong impact on the discrepancy since it was shown early on by \citet{2014A&A...571A..29P} that adopting a different value for the mass observable parameter could alleviate the tension and that adopting a  different value for  the scaling calibration  may serve as a solution to fix the discrepancy.
Therefore, several studies investigated the impact of modifications to the way the mass observable relations are determined on the cluster abundance observable with relative success when trying to account for it as solution to the $\sigma_8$ discrepancy. One line targeted the cooling and  heating processes closely related to cluster mass determination, such as the work by \cite{SPT:2017ees}, \cite{Henson:2016eip}, and  \cite{Mummery:2017lcn}, who showed, using hydrodynamic simulations, that using a different mass weighted cooling scheme or taking into account non-thermal pressure coming from bulk and turbulent motions of gas in the intracluster medium \citep{Shi:2015fua,Ettori:2021bug} could reduce the hydrostatic bias, although still below the amount needed to reconcile the observed cluster number count with CMB. Another study was performed by the second release of the Planck collaboration \citep{2016A&A...594A..27P} on the impact of the signal-to-noise ratio (S/N)   on the number of detections, but found that it would be necessary for Planck to have missed nearly 40\%\ of the clusters with predicted SZ S/N > 7 in order for the SZ
and CMB cluster number count to match with a level of scatter with cluster structure inconsistent with the predictions of current hydrodynamical simulations. More recently, \cite{Lovisari:2020ynb} investigated the differences in the mass estimates obtained through independent X-ray, weak-lensing, and dynamical studies in the Planck-ESZ  cluster sample and found an average value insufficient to reconcile the Planck cluster abundance and CMB results. Using simulations, \cite{Castro:2020yes} and \cite{Debackere:2021ado} studied  the effect of baryonic physics on cluster statistics and found it subdominant  compared to other systematics for current available observations, similarly to \cite{Bocquet:2015pva}. \cite{Salvati:2021gkt} showed that we need to drop the self-similarity hypothesis and assume a change of  $\sim$ 4 $\sigma$ in the slope of the scaling relation to reconcile CMB and cluster counts similar to \cite{Andreon:2014fea}, while \cite{Ghirardini:2020vyv} and \cite{Wicker:2022ahi} investigated  a possible redshift evolution of the mass observable bias and found a value for the amplitude of the bias that is consistent with that expected from hydrodynamical simulations and local measurement, similarly to \cite{Salvati:2019zdp}.\\ 
Another source of misdetermination of the cluster mass comes from possible overlooked biases from the weak lensing proxy. \cite{Rasia:2012jz} showed that the scatter in the mass lensing relation could be reduced
by discarding clusters having an irregular morphology or clusters influenced by substructure-rich
environment.  Another bias according to \cite{Valageas:2013qfa} could arise from   galaxies
that are not distributed randomly in space, but are correlated with the underlying density field, resulting in a source-lens clustering. \cite{Bahe:2011cb} and \cite{Hamana:2012ur} found that the standard method for identifying clusters  through a search of the peak heights in the lensing convergence map is not well suited to determining the cluster mass as the peaks are poorly
correlated with their virial mass. More recently, \cite{Meneghetti:2020yif} reported that the observed cluster substructures are more
efficient lenses than predicted by the  CDM simulation. This could be the reason why (see the study by \cite{Lovisari:2020ynb})  the mean mass ratios inferred from the Weak Lensing (WL) masses of different projects vary by a large amount, from the APEX-SZ survey \cite{Klein:2019dru} showing a very small bias, to the Local Cluster Sub-structure Survey (LoCuSS) \citep{Zhang:2008tr} and the combined Multi Epoch Nearby Cluster Survey (MENeaCS) with the Canadian  Cluster Comparison Project (CCCP) \cite{Herbonnet:2019byy} showing a more significant value ($1-b \sim$ 0.77), to the aforementioned WtG \citep{2014MNRAS.439...48A} pointing to the largest deviation ($1-b \sim$ 0.6), a value that would alleviate the tension between the Planck results. The Planck SZ collaboration also elaborated a novel method for measuring cluster masses based on matching lensing of the CMB temperature anisotropies behind clusters \citep{Melin:2014uaa} and obtained $1-b \sim 1.01 \pm0.2$, a value that increases the discrepancies although with greater statistical uncertainty. More recently, mixed results have been found by \cite{Ingoglia:2022fjk} and \cite{Lesci:2020qpk} when using lensing to calibrate the mass of a sample of clusters from the third KiDS release;  the first is in agreement with the Planck's clusters calibration, while the second is consistent with CMB inferred counts. \cite{Murray:2022iyi} presented weak lensing mass estimates for a sample of 458 galaxy clusters from the redMaPPer
Sloan Digital Sky Survey and found a tension between the mass--richness relation and that from DES collaboration cluster sample. Finally, \cite{Schrabback:2020hxx} performed weak lensing mass calibration on distant SPT galaxy clusters from expanded follow-up deep observations and found it 30\% lower than needed to reconcile a Planck cosmology with the observed SPT-SZ cluster counts.\\
Other halo properties than the halo mass that could impact the abundance of clusters should also not be forgotten, such as formation time,   concentration, and   the dependence of the spatial distribution of dark matter halos, all commonly called halo assembly bias (see \cite{Sunayama:2015zqa}, and reference therein), but also the cluster triaxiality or shape and orientation impact on the number of detections (see \cite{Zhang:2022urs} for a recent assessment on the DES detection of clusters) or other environmental effects, such as the density of the surrounding background  \citep{Manolopoulou:2020vxf}, and the impact on X-ray clusters.\\
To summarise, almost all cluster-based probes show a discrepancy with the  CMB probe on the $\sigma_8$ constraints when using their mass-observable calibration obtained by different methods  (see  \cite{2017A&A...604A..89P} and \cite{2018A&A...614A..13S}, for a compilation) and only few (some of those mentioned above) suggest that there are improper selection functions, due to a neglected Eddington bias \citep{2016JCAP...08..013B} or a Malmquist bias \citep{2017MNRAS.464.2270A} for example, that should be taking into consideration and could result in a substantial reduction of the discrepancy. Consequently, this has motivated us to consider other alternatives as possible origins of this tension.

Among these alternatives, one was an extension to the $\Lambda$CDM model by adding massive neutrinos. The latter escape matter potentials on scales below their free streaming distance, slowing down the growth of matter perturbations \citep{2010JCAP...09..014B} and lowering the value of $\sigma_8$. Therefore, several studies, with more or less favourable conclusions, tried to test whether a non-minimal neutrino mass helps in reducing the gap between the two constrained values of $\sigma_8$ \citep{2014PhRvD..90h3503D,2014PhRvD..90d3507G,2014PhRvD..90h3503D,2016PhLB..752..182D}.  
In the companion papers \citet[][hereafter SK18,]{2018A&A...620A..78S} and \citet[][hereafter IL19]{2019A&A...631A..96I}, we also examined  with Markov chain Monte Carlo  (MCMC) techniques the role of neutrinos in fixing this discrepancy using an X-ray cluster sample or SZ detected cluster sample, in combination with CMB datasets and additional secondary probes like BAO and Ly-$\alpha$, and found that massive neutrinos are not able to alleviate the tension because  the introduction of massive neutrinos modifies the confidence contours on $\sigma_8$, from cluster counts and those from the CMB data, in the same direction of the $\sigma_8 - \Omega_m$ degeneracy, thus does not reduce the gap in the value of $\sigma_8$ obtained from the two probes.
We  explored  another possible origin for this discrepancy through a modification of general relativity. This could result in a change in the growth rate of structures, which is   parametrised in function of the so-called growth index $\gamma$ \citep{1980lssu.book.....P}. Different values of $\gamma$, corresponding to different modified gravity models,   change the expected number of clusters, yielding different constraints on $\sigma_8$. Therefore, in \citepalias{2018A&A...620A..78S} and \citepalias{2019A&A...631A..96I}, we investigated the change in $\sigma_8$ constraints when allowing a free $\gamma$ in the presence of massive neutrinos. However, unlike other works, we left the mass observable scaling factor free to vary when combining the two aforementioned probes. This data-driven approach was motivated by \cite{2005A&A...436..411B}, where $\sigma_8$ constraints were found correlated with those of the mass calibration parameter. Therefore, any new parameter showing correlation with the mass observable one would be able, within the allowed constraints, to fix the discrepancy. In  \citetalias{2018A&A...620A..78S}, we found that $\gamma$ satisfies this condition, independently from the value of the neutrino mass, thus indicating its ability, unlike neutrinos, to fix the discrepancy. However, this is only possible for values far from the fiducial $\Lambda$CDM values that are not in agreement with constraints from other growth probes. Moreover, we also found in \citepalias{2019A&A...631A..96I}, when using instead the deeper SZ cluster detected sample that although $\gamma$ still correlates with the calibration parameter,  it cannot reach the value necessary to fix the discrepancy, due to the constraining power of the SZ sample probing a deeper and wider redshift range. 
\\
Drawing from the same idea, we propose here to explore the impact on the aforementioned discrepancy of another extension to $\Lambda$CDM through a change in the value of the dark energy equation of state (EoS) parameter $w$. One way of doing this is by allowing a different constant value for $w$ than the $\Lambda$CDM equivalent $-1$. Another would be through the phenomenological relation $w=w_0 +(1-a) \, w_a$ \citep{2001IJMPD..10..213C,2003PhRvL..90i1301L} mimicking a redshift dependent scalar field or other equivalent model capable of inducing similar changes in dark energy density. There were studies investigating the effect of a variation of EoS on the constraints of the cosmological parameters, especially $\sigma_8$, using cluster counts or probes at the local universe sensitive to the growth of structures. However, unlike the method we followed in the companion papers of this work, they were all mostly done after fixing the cluster observable mass calibration to that determined from the hydrodynamical simulations. Among these works, \cite{2017MNRAS.469.1713Z} used CMB $C_{\ell}$ combined with SZ cluster counts to put constraints on the dark energy EoS parameter $w$ and neutrinos mass when the two are left free to vary. They found that a $w$ different from its fiducial value, is favoured by this combination and impacts the weighing of neutrinos, thus indirectly suggesting that a $w$, alone or in combination with neutrinos, could be a solution to the discrepancy.    
In addition, \cite{2017MNRAS.471.1259J} compared the $\sigma_8$ constraints from local weak gravitational lensing and that inferred from the CMB $C_{\ell}$ and found that a value for the dark  energy EoS parameter far from the equivalent fiducial $\Lambda$CDM value $w=-1$ could alleviate the tension. Recently, \cite{2018MNRAS.477.4957B,2018A&A...614A..13S} and \cite{2019ApJ...878...55B}, using another galaxy cluster-based probe, the galaxy cluster power spectrum in combination with CMB $C_{\ell}$, showed that a $w \neq -1$ is favoured over $\Lambda$CDM. 
Here we also investigate whether the variation of the  dark energy parameter, alone or combined with neutrinos, could significantly reduce the cluster counts discrepancy and alleviate the tension. 
 Furthermore, we test how the $\sigma_8$ - $\Omega_{\rm m}$ degeneracy behaves when we allow the growth index to vary along with the dark energy EoS parameter. We perform this study, using a SZ cluster sample in combination with CMB Planck $C_{\ell}$ as our main datasets in a MCMC parameter inference analysis. In line with our previous investigations, we leave the cosmological parameters of the standard model, the nuisance parameters, and the calibration parameter free to vary in a model independent approach, letting the data fix the last parameter.
 To alleviate possible degeneracies, we also consider an additional constraining probe, the BAO feature angular distance, which is more sensitive to certain expansion parameters (e.g.  the Hubble constant) or to certain redshift ranges than the others. 
 \\
 This study is organized in the following way.  In Sect. \ref{sec:ThC} we review briefly the theoretical determination of the cluster counts observable within the extensions we consider:  the dark energy density EoS parameter, the growth index, and the neutrino mass. In Sect. \ref{sec:impL} we describe the datasets we used and how we implemented them in the MCMC analysis. We then present the results obtained in Sect. \ref{sec:resuL} along with further evidence tests in Sect. \ref{sec:furthertest} and conclude in Sect. \ref{sec:concL}.\\

\section{Cluster abundance and extensions to $\Lambda$CDM}
\label{sec:ThC}
Cluster abundance formalism and theoretical calculations in the $\Lambda$CDM model or extensions to it, along with sample construction and selection functions for the SZ detected cluster samples, is described in \citet{2016A&A...594A..13P}, \citetalias{2018A&A...620A..78S}, and \citetalias{2019A&A...631A..96I}.
Here we briefly summarise most of the theory and methods used when massive neutrinos or modified gravity are introduced, along with a dark energy EoS parameter different from the fiducial $\Lambda$CDM.

\subsection{Mass function and cluster counts}

Under the general hypothesis of self-similarity, the halo mass function (HMF) can be written in a simple form \citep{1992A&A...264..365B} :
\begin{equation}\label{eq:nm0}
  n(m)=-\frac{\bar{\rho}}{m}\frac{\rmd{\rm \nu}}{\rmd m} \mathcal{ F}({\rm \nu}) 
.\end{equation}
Here $\rho_0$ is the mean matter density today and $\mathcal{ F}({\rm \nu})$ is a functional form of ${\rm \nu} = \delta_{c}/\sigma(m)$, the normalised amplitude of fluctuations, where $\delta_c$ is the linear density threshold at the present time for non-linear collapse and $\sigma(M)$ is the rms of the linear density perturbations within a sphere of radius $R$ that contains mass $M=4\pi/3\,\rho_0 R^3$. The original form of the function $\mathcal{F}$ derived by \cite{1974ApJ...187..425P} is
\begin{equation}\label{eq:PS-MF}
  \mathcal{ F}_{\rm PS}({\rm \nu}) =\sqrt{\frac{2}{\pi}}\exp\left[-\frac{{\rm \nu}^2}{2}\right].
\end{equation}
A more refined evaluation of $\mathcal{F}$ has been the subject of numerous investigations and 
continuous improvements of simulations that has allowed a more accurate evaluation of the mass function such as \cite{2016MNRAS.456.2486D} (DP16) HMF we use in this work, which updated the original evalutation of \cite{1999MNRAS.308..119S},  
\begin{equation}
    \mathcal{ F}_{\rm D}(\nu) =A\sqrt{\frac{2a}{\pi}}\left[1+\left(\frac{1}{a\nu^2}\right)^p\right]\  \ \exp\left[-\frac{a\nu^2}{2}\right]
,\end{equation}
with new values for the fitted coefficients $A$, $a$, and $p$. 
Then, in order to compare the theoretical HMF to the actual measured abundance of clusters of galaxies, we need a relation between the cluster mass entering the mass function and its observable quantities. This is done in the present study using 
a sample of  SZ detected clusters,  where the distribution of clusters function of redshift and S/N is written as
\begin{equation}
\frac{dN}{dz dq} = \int d\Omega_{\rm mask} \int d{M_{500}} \, \frac{dN}{dz d{M_{500}} d\Omega}\, P[q | {\bar{q}_{\rm m}}({M_{500}},z,l,b)],
\end{equation}
with
\begin{equation}
\label{eq:dndzdq}
\frac{dN}{dz d{M_{500}} d\Omega} = \frac{dN}{dV d{M_{500}}}\frac{dV}{dzd\Omega},
\end{equation}
and the quantity $P[q | {\bar{q}_{\rm m}}({M_{500}},z,l,b)]$ being the distribution of $q$ given the mean S/N value, ${\bar{q}_{\rm m}}({M_{500}},z,l,b)$, predicted by the model for a cluster of mass ${M_{500}}$ (i.e. the overdensity defined at 500  with respect to the critical density of the universe) and redshift $z$ located at Galactic coordinates $(l,b)$.\\

We relate this quantity to the measured integrated Compton $y$-profile $\bar{Y}_{500}$ using the  scaling relations 
\begin{equation}
\label{eq:Yscaling} 
E^{-\beta}(z)\left[\frac{{D_{\rm A}}^2(z) {\bar{Y}_{500}}}{\mathrm{10^{-4}\,Mpc^2}}\right] =  Y_\ast \left[ {\frac{h}{0.7}}
  \right]^{-2+\alpha} \left[\frac{(1-b)\,
    {M_{500}}}{6\times10^{14}\,M_{\odot}}\right]^{\alpha},
\end{equation}where ${D_{\rm A}}$ is the angular diameter distance and
$E(z) = H(z)/H_0$, while  $\alpha$, $\beta$, and $Y_\ast$ are additional parameters in the SZ scaling law, along with $(1-b)$, which serves to link  $M_{500}$ to $M_{\rm X}$, the cluster mass determined from X-ray observations, playing the role of the calibration parameter obtained from comparison with hydrostatic simulations. There is  an additional scaling parameter $\sigma_{\rm{ln}\,Y}$ that corresponds to the scatter of the log-normal distribution that the measured $Y_{500}$ is assumed to follow with $\bar{Y}_{500}$ the mean, which we also take into account in our analysis.\\

\subsection{Extensions to the $\Lambda$CDM model and their implementation}

\subsubsection{Dark energy}
\label{sect:DEcpl}

A possible origin behind the discrepancy on $\sigma_8$ is that between the time of recombination and the present time, a deviation from the $\Lambda$CDM equivalent dark energy fluid EoS value changes the growth of structure in such a way that the $\sigma_8$ is modified to the desired quantity.\\
The most common way to describe this deviation is by choosing a dark energy EoS parameter $w$ constant $\neq-1$ as a first simple extension of $\Lambda$CDM, the $w$CDM. We could also consider further expansions of the dark energy EoS through the CPL phenomenological parametrisation mentioned in Sect.~\ref{sec:IntrO}: 
\begin{equation} 
 w=w_0 +(1-a)w_a
 \label{eq:CPL}
.\end{equation}
  This will modify the dark energy density $\rho_{\Lambda}$ entering $\Omega_\Lambda$ in the Friedmann-Lema\^{\i}tre equations to 
\begin{equation}\label{equ:rhoDE}
\rho_{\Lambda\left(mod\right)}=\rho_{\Lambda}\left(1+z\right)^{3\left(1+\omega_{0}+\omega_{a}\right)}\exp\left(-3\omega_{a}\dfrac{z}{z+1}\right)
,\end{equation}
with influence on the equation governing the growth of structure,
\begin{equation}\label{perTgroW}
\delta''(a)+\left[\frac{3}{a}+\frac{H'(a)}{H(a)}\right]\delta'(a)-\frac{3}{2}\Omega_{\rm m}(a)\frac{H_0^2}{H^2(a)}\frac{\delta(a)}{a^5}=0\,,
\end{equation}
where a $w$ constant or redshift dependent enters the expression of $H(a)$ and $\Omega_{\rm m}(a)$.\\
In addition to  accounting for $w$ in the density budget and the growth rate in the halo mass function, we leave the value of $\delta_c$ equal to its fiducial value $\sim 1.686$, since \citet{Horellou2005} and \citet{Pace2010} have shown that $\delta_c$ remains almost unchanged when considering dark energy with a $w\neq-1$ that is  constant or evolving with redshift.\\
For the HMF multiplicity, several studies \cite[see][and reference therein]{2012PDU.....1..162B} showed that the standard fitting performed in the  $\Lambda$CDM framework still provides a good fit to the  results from  the simulations with  constant $w\neq$-1 and parametrised EoS dark energy cosmologies, providing we correct the growth rate of linear density perturbations to the value induced by the background scaling of the dark energy density, as we do here.
 Therefore, we can still assume the same functional HMF and its redshift evolution provided by   DP16 :
\begin{align*}
 a = &\, 0.4332x^2 + 0.2263x + 0.7665\,,\\
 p = &\, -0.1151x^2 + 0.2554x + 0.2488\,,\\
 A = &\, -0.1362 x + 0.3292\,.
\end{align*}
Here $x=\log{(\Delta_{500}(z)/\Delta_{\rm{Vir}}(z))}$
is used to convert from the virial mass definition to the critical cluster mass definition the SZ sample selection function is built upon.\\
Finally, we rely on \citet{2009A&A...496..637A}, who showed that considering dark energy cosmologies does not significantly alter the standard galaxy cluster scaling relations,  to keep the same functional form for Eqs.~\ref{eq:dndzdq} and ~\ref{eq:Yscaling}, which  are therefore kept when varying the EoS parameter $w$.

\subsubsection{Growth index}

The growth rate $ f=d\ln \rm D / d\ln  a$,  where $D$ is the growth of perturbations $\delta(z) = \delta_0  D(z)$, could be approximated by $ f=\Omega_{m}^\gamma(z)$ 
with $\Omega_{ m}(z)$ the matter density at $z$ and $\gamma$, called the growth index. The growth rate in $\Lambda$CDM is well approximated when the growth index is set to $\sim 0.545$ \citep{1990ApJS...74..831L,1991MNRAS.251..128L} and takes different values in other modified gravity models \citep{2005PhRvD..72d3529L}. Since late large-scale structures growth could influence the cluster number counts, and thus impact the $\sigma_8$ values, a solution to the discrepancy from allowing a free growth index was investigated in \citetalias{2018A&A...620A..78S}. \\
However,  since   the effect of $\gamma$ on the growth rate could also be degenerate with those of $w$, we investigate in this work how the constraints on the cosmological parameters, in particular $\Omega_{\rm m}$ and $\sigma_8$, behave when we allow both $\gamma$ and $w$ the dark energy parameter to vary, following~

\begin{equation}\label{eq:sigmodgam}
    \sigma_{\gamma,wCDM}(z)=\frac{D_{\Lambda CDM}(z_*)}{D_{\gamma,CDM}(z_*)}\frac{D_{\gamma,CDM}(z)}{D_{\Lambda CDM}(z)}\,\sigma_{wCDM}(z)
.\end{equation}In addition, we consider a redshift dependent growth index following a recent higher order parametrisation proposed by \cite{2014JCAP...05..042S}: 
\begin{equation}
\label{eq:gamMar}
\gamma= \gamma_0 + \gamma_1 \ln \left(\Omega_{m}\left(z\right)\right)
.\end{equation}

\subsubsection{Massive neutrinos}

Neutrinos, when massive, contribute to the matter density as part of the cosmological energy density budget with implications on the CMB spectrum. However, although they are part of the matter density at late time, they do not contribute to the growth of the linear matter power spectrum, an ingredient of the cluster counts probe, but rather reduce it by a factor $(1-f_{\rm \nu})^2$, where $f_{\rm \nu}\equiv\Omega_{\rm \nu}/\Omega_{\rm m}$ \citep{2006PhR...429..307L}.\\
There is  an additional  non-linear effect from massive neutrinos that needs to be taken into account when evaluating clusters counts. Simulations by \citet{2013JCAP...12..012C} and \citet{2014JCAP...02..049C} have shown that we get a better match between a simulated and calculated mass function if we use what is  known as the CDM prescription, where the matter density in the calculation is replaced with the cold dark matter density and the power spectrum is replaced by the cold dark matter power spectrum.\\
Similar to the companion papers \citepalias{2018A&A...620A..78S}, in order to investigate if the above effects from neutrinos could mix with those of the aforementioned $\Lambda$CDM extensions and influence their ability to fix the discrepancy, we  further consider here the case with a free neutrino mass in addition to a free dark energy EoS parameter and a free growth index $\gamma$.\\
Finally, massive neutrinos are considered in the standard scenario of particle physics where the effective number of relativistic particles $N_{\rm eff}$ is equal to $3.046$. However,  beyond the standard model, the physics may alter this value due to the presence of extra relativistic contributions. Therefore, in Appendix \ref{effNeff}, the effect of varying $N_{\rm eff}$ with $w\neq-1$ is also discussed. 

\section{Datasets and methods}
\label{sec:impL}

As one of our primary probes, we use the CMB $C_{\ell}$ of the  temperature, polarisation, and their  cross-correlations from the publicly available datasets of the Planck mission \citep{Planck:2019nip} and their publicly provided likelihood. We combine them with the SZ detected cluster sample PSZ2 containing a total of 439 clusters \citep{2016A&A...594A..27P} spanning the redshift range from z$\sim$0.0 to z$\sim$1.3. We  further constrain the dark energy EoS using the scale of the baryonic acoustic oscillation (BAO)  of the galaxy distribution measurements  \citep{2011MNRAS.416.3017B,2015MNRAS.449..835R,2017MNRAS.470.2617A}. The inference of the constraints on the parameters is done using standard MCMC techniques where observations are confronted with theoretical angular power spectra calculated with the Boltzmann solver \texttt{CLASS} \citep{2011arXiv1104.2932L}. \texttt{CLASS}  allows  the effects of different neutrino masses and species to be added, as well as different  dark energy evolution scenarios.
The total neutrino mass $\sum m_{\rm \nu}$ varies in the most general configuration with three degenerate massive active neutrinos. Later the  $N_{\rm eff}$ parameter is also left free to vary. Then using the scaling relation of  Eq.~\ref{eq:Yscaling} with the calibration parameter $(1-b)$  for the  SZ clusters, we relate the mass to the  SZ signal to construct the redshift bins cluster counts. All is implemented in our SZ cluster counts module in the framework of the parameter inference Monte Python code \citep{2013JCAP...02..001A}.
When running MCMC chains, we mainly  let  the normalisation parameter $(1-b)$ for SZ free to vary along with the six CMB cosmological parameters. For the other cluster scaling parameters, we fix $\beta$ and $Y_\ast$ to their best values obtained from simulations $\beta \sim 0.66$ and log$Y_\ast = -0.186$, while leaving $\alpha$ free. We also leave $\sigma_{\rm{ln}\,Y}$ free with Gaussian priors derived from its best fit $(= 0.075 \pm 0.01)$.\\
We note that we do not take into account cross-correlations between different experiments as they are all, except the CMB measurements, almost independent and do not overlap in time or space with each others. Later in this work, we  need to compute the Bayesian evidence (with respect to the model with the adopted Planck collaboration calibration) of the alternative extensions we take into consideration for addressing the $\sigma_8$ tension. Considering a dataset $\mathbf{x}$ and two different models ${\cal M}_i$ and ${\cal M}_j$, described by the parameters $\boldsymbol{\theta}_i$ and $\boldsymbol{\theta}_j$. The Bayes factor of model ${\cal M}_i$ with respect to model ${\cal M}_j$, denoted $B_{ij}$, is given by
\begin{eqnarray}
B_{ij} \equiv \frac{\int d\boldsymbol{\theta}_i\, \pi(\boldsymbol{\theta}_i \vert {\cal M}_i) {\cal L}(\mathbf{x} \vert \boldsymbol{\theta}_i,{\cal M}_i)\,,}{\int d\boldsymbol{\theta}_j\, \pi(\boldsymbol{\theta}_j \vert {\cal M}_j) {\cal L}(\mathbf{x} \vert \boldsymbol{\theta}_j,{\cal M}_j)\,,}\,,
\label{eq:bayesfactor}
\end{eqnarray}
where $\pi(\boldsymbol{\theta}_i \vert {\cal M}_i)$ is the prior for the parameter $\boldsymbol{\theta}_i$, and ${\cal L}(\mathbf{x} \vert \boldsymbol{\theta}_i,{\cal M}_i)$ is the likelihood of the data given the model parameter $\boldsymbol{\theta}_i$. A Bayes factor $B_{ij}>1$ (or equivalently $\ln B_{ij}>0$) indicates that model ${\cal M}_i$ is more strongly supported by data than model ${\cal M}_j$.\\
Finally, for the purpose of quantifying the degree of correlation between the parameters inferred from MCMC, we later calculate the \citeauthor{1895RSPS...58..240P} coefficient $r_{xy}$ for a set of paired values of the parameters we want to investigate, following the formula  
\begin{equation}
r_{xy} =\frac{n\sum x_iy_i-\sum x_i\sum y_i}
{\sqrt{n\sum x_i^2-(\sum x_i)^2}~\sqrt{n\sum y_i^2-(\sum y_i)^2}}
,\end{equation}
where $n$ is the sample size and $x_i$, $y_i$ are the individual sample points indexed with $i$. The coefficient has a value between +1 and -1, where 1 is total positive linear correlation, 0 is no linear correlation, and -1 is total negative linear correlation.


\section{Results}
\label{sec:resuL}
In \cite{2014A&A...571A..29P} and \cite{2015A&A...582A..79I}, it
was shown that the discrepancy in galaxy cluster counts, could translate into that of the mass observable calibration parameter. In particular, in a MCMC analysis combining CMB and cluster counts with a free calibration parameter, the  preferred value for the free parameter was found to be $(1-b)\sim 0.6$, corresponding to the $\sigma_8$ obtained from Planck CMB datasets, while the $(1-b)$ calibration value adopted for the Planck SZ cluster sample  is equal to $0.8$ more than 4 $\sigma $ away from the above best value. Massive neutrinos were advocated as a possible solution to alleviate the tension. However, it was found in \citetalias{2018A&A...620A..78S} and \citetalias{2019A&A...631A..96I} that the combination of CMB $C_{\ell}$ and cluster counts still yield the same constraints on the calibration parameter, independently from neutrino mass, indicating that the latter is not able to fix the discrepancy. Another extension to $\Lambda$CDM by means of a growth index $\gamma$ different from the fiducial $\Lambda$CDM value of $\sim 0.55$ was also investigated as a way to alleviate the tension. Following the same method of leaving the calibration parameter free to vary when combining CMB $C_{\ell}$ and cluster counts, a correlation between the calibration parameter and the growth index was observed, suggesting that the latter could alleviate the tension. However, this was only when using an X-ray cluster sample at very low redshift, while the constraints from the SZ deeper Planck cluster sample \citepalias{2019A&A...631A..96I} are forbidding $\gamma$ from reaching values that would allow $(1-b) \sim 0.8$. Here we follow the same approach as in \citetalias{2018A&A...620A..78S} and \citetalias{2019A&A...631A..96I}, specifically   we combine the CMB angular power spectrum and cluster counts probes in a MCMC analysis to constrain $\sigma_8$ and the other cosmological parameters, leaving the mass calibration parameter free in order to be fixed 
by data and allowing, instead of $\gamma$, a free dark energy EoS parameter $w$ as our extension to $\Lambda$CDM. We also explore other cases where we additionally allow $\gamma$ or neutrinos to vary alongside $w$ to investigate the impact of these combinations on the discrepancy. We limit our study to the SZ cluster sample only, since the X-ray sample is much less constraining, and it was shown in \citetalias{2019A&A...631A..96I} that the combination of the two samples does not have any significant influence on the constraints obtained from the SZ alone.

\subsection{Dark energy equation of state parameter and the $\sigma_8$ discrepancy}\label{subsec:addw0}

\begin{figure}[ht]
\centering
\includegraphics[width=\columnwidth]{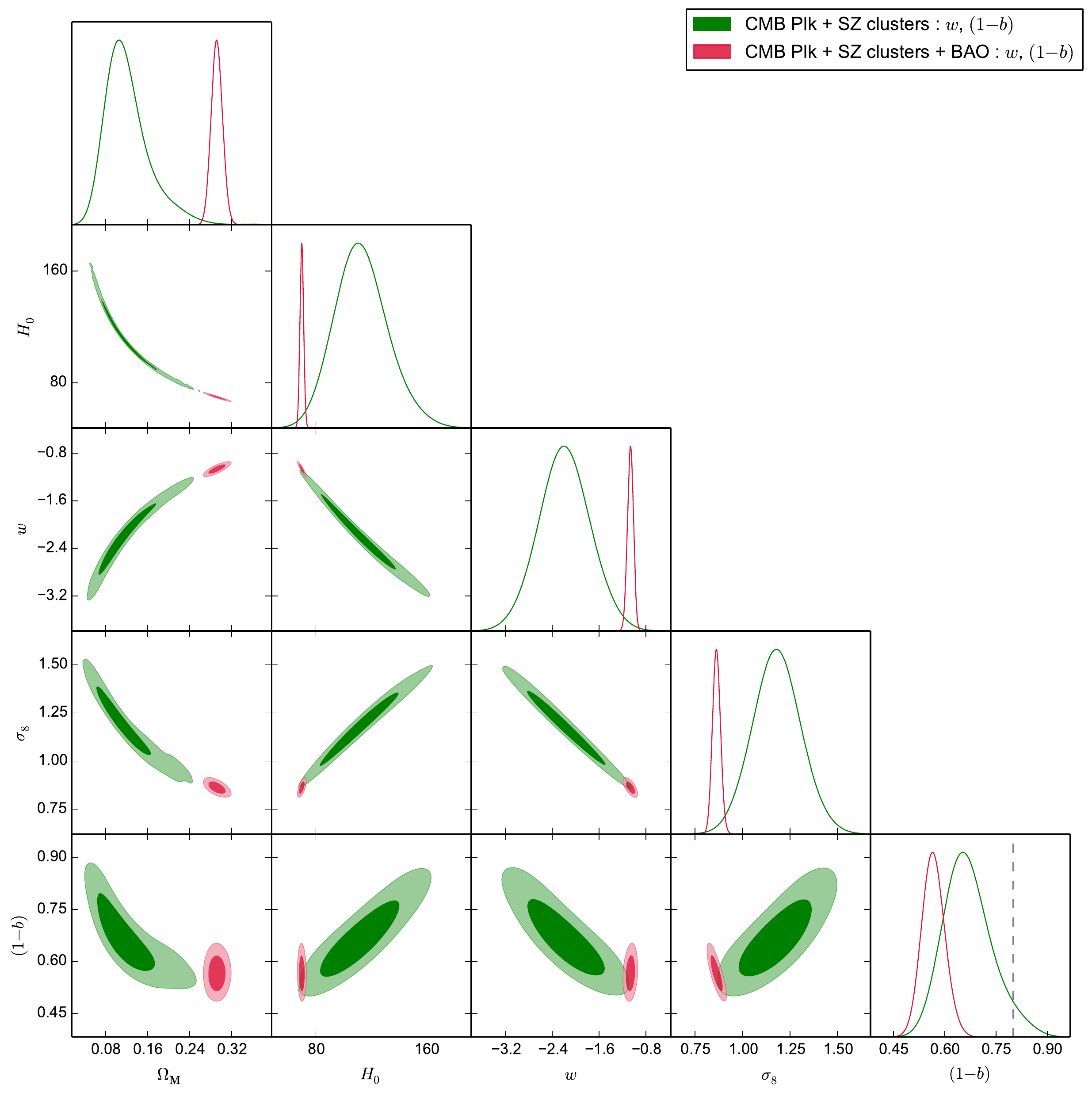}
\caption{Two-dimensional  confidence contours at 68\% and 95\% for the parameters $\Omega_{\rm m}$, $H_0$, $\sigma_8$, $1-b$, and $w$ derived from Planck 2018 CMB datasets combined with the Planck 2015 SZ detected galaxy cluster sample (green). Adding the  BAO probe datasets, still with a free constant $w$,  tightens the constraints considerably.  The vertical black dashed line gives the Planck calibration value $\left( 1-b \sim 0.8 \right)$.}
\label{fig:plkjlaSZATM3nuwo}
\end{figure}

In Fig.~\ref{fig:plkjlaSZATM3nuwo} we show confidence contours for the parameters $\sigma_8$ and $\Omega_m$ along with $(1-b)$, the mass observable calibration parameter, obtained from a MCMC analysis using CMB Planck datasets in combination with the cluster counts from the  Planck SZ detected sample, while allowing $w$, the dark energy equation parameter, to vary. We also perform the same analysis, but by  combining it with BAO. In the first case (green contours and lines) the preferred calibration is still low: $1-b\sim 0.64$;     the tension with $1-b\sim 0.8$ is largely reduced, being  less than 2$\sigma$. However, with $w$ tightly correlated with $(1-b)$, the agreement happens for values of $w \sim -3$ far below $-1$, pushing cosmological parameters in a region:  $ H_0 \sim 140$ km/s/Mpc and  $ \Omega_m \sim 0.08$ largely excluded by other cosmological probes.  With the addition of the BAO data  (red contours and lines) the   $w$ variation is now limited to  a small region around $w \sim -1$, while the  $(1-b)$ preferred value returns to that of $\Lambda$CDM $\sim 0.6$ with Planck SZ cluster  calibration at more than 5$\sigma$ away.

\subsection{Dark energy and growth index interplay}
\label{subsec:modeltmpfct}

\begin{figure}[ht]
\centering
\includegraphics[width=\columnwidth]{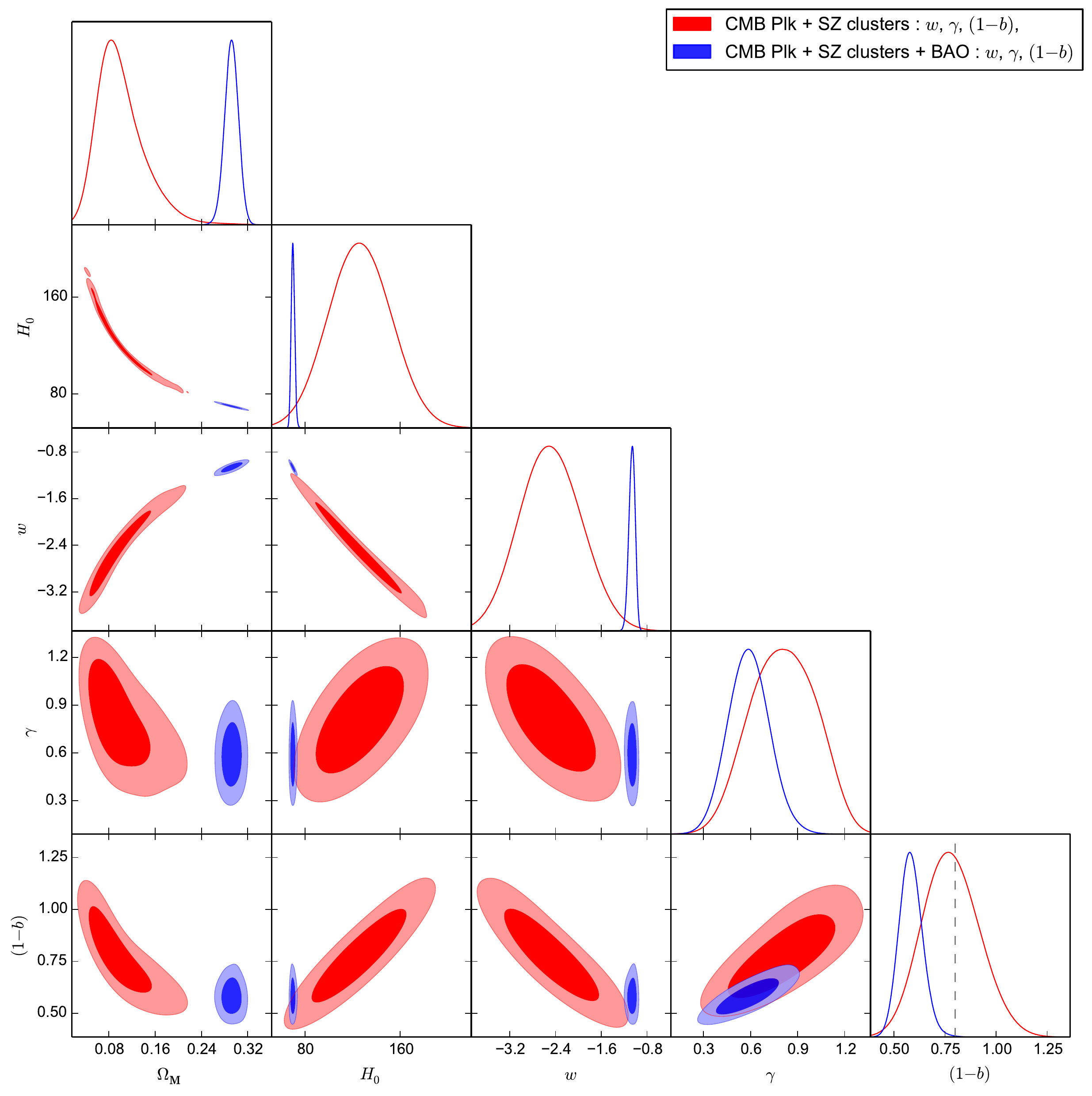}
\caption{Confidence contours at 68\%\ and 95\%  for the parameters $(1-b)$, $\sigma_8$, $\Omega_{\rm m}$, $H_0$, $w$, and $\gamma$ derived from    Planck 2018 CMB datasets combined with the Planck 2015 SZ detected galaxy cluster sample, then adding  BAO probe datasets, all with a free constant $w$ and $\gamma$. The vertical black dashed line gives the Planck calibration value $\left( 1-b \sim 0.8 \right)$.}  
\label{fig:plkSZgam3w0}
\end{figure}

As seen above, the dark energy EoS parameter is not able to reduce the discrepancy to less than 2$\sigma$. This is mainly limited by the need of $w$ to accommodate  both volume change with cluster sample redshift bins and the variation in the growth rate in the sample bins. This was also more notable in \citetalias{2019A&A...631A..96I} when we introduced a growth index $\gamma$ to alleviate the tension, without success despite observing a strong correlation between $\gamma$ and the calibration factor $(1-b)$. To see if the combination with the SZ sample is able to simultaneously constrain two degrees of freedom when we allow  $w$ and $\gamma$ to vary simultaneously, we show in Fig.~\ref{fig:plkSZgam3w0} the results of a MCMC analysis combining CMB datasets and SZ cluster counts, letting all the cosmological parameters free to vary along with the growth index $\gamma$, $w$, and the mass-observable normalisation factor. We also perform the same analysis adding BAO probes. We observe in the first case (red lines and contours) that the combination of $w$ with $\gamma$ succeeds in alleviating the tension with the $(1-b)$ preferred value, clearly shifting to become  centred  around $0.8$.  However, this is obtained with the other cosmological parameters shifting far away from their standard values, so that when we add the BAO probes (blue lines and contours), we again obtain  strong constraints on all parameters around the fiducial $\Lambda$CDM values, and the Planck SZ calibration value (dashed black vertical line) becomes barely reachable at $3\sigma$ level. 

\begin{figure}[ht]
\centering
\includegraphics[width=\columnwidth]{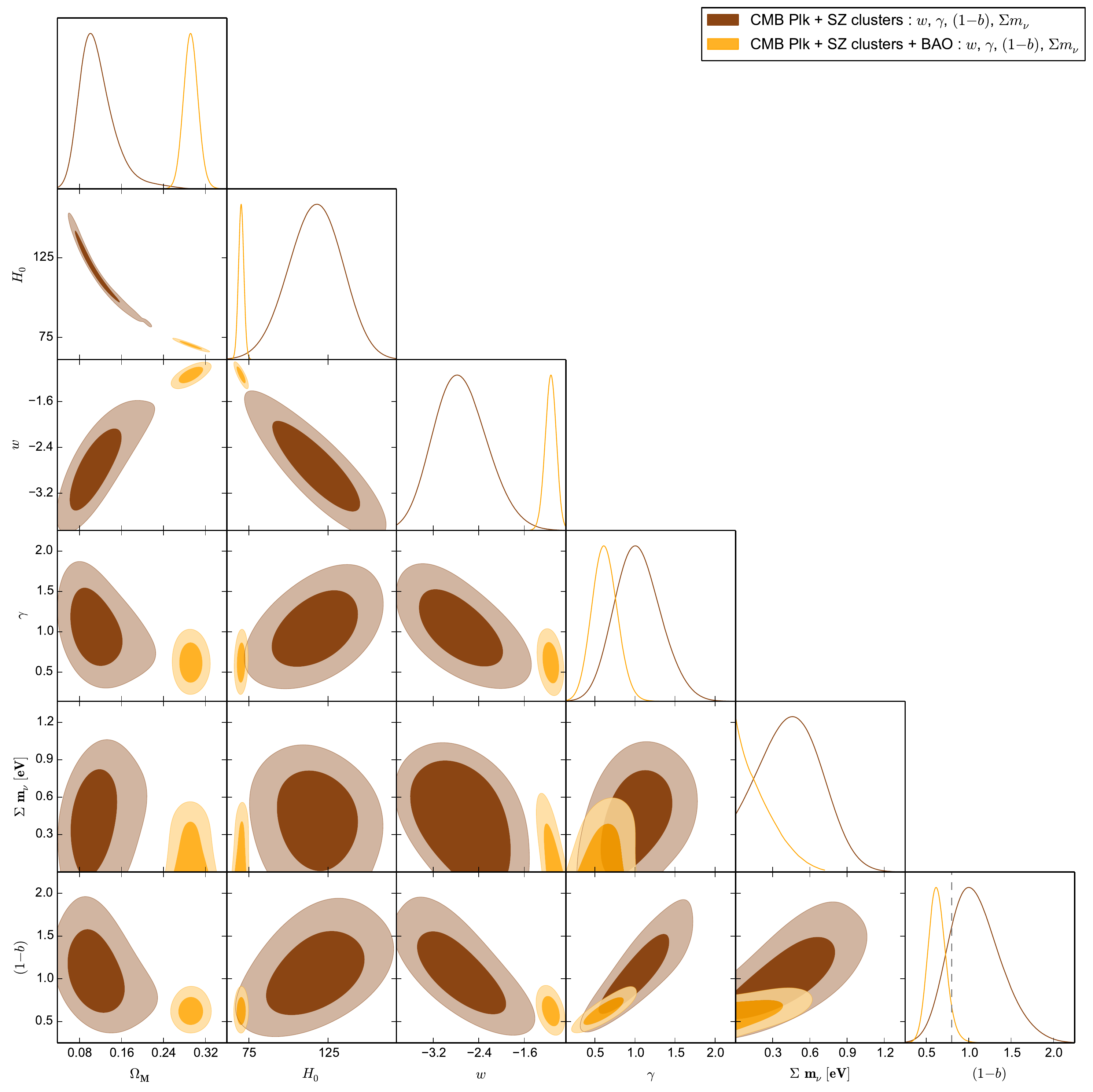}
\caption{Confidence contours at 68\%\ and 95\%  for the parameters $(1-b)$ , $\sigma_8$, $\Omega_{\rm m}$, $H_0$, $w$, $\gamma$, and $\Sigma m_{{\rm \nu}}$ derived from Planck 2018 CMB datasets combined with  the  Planck 2015 SZ detected galaxy cluster sample, then adding BAO probe datasets, all with a free constant $w$, $\gamma$, and neutrino mass. The vertical black dash line gives the Planck calibration value $\left (1-b \sim 0.8 \right )$.}
\label{fig:plkSZgam3nuw0}
\end{figure}

We also show the case when we allow the mass of neutrinos to vary  in Fig.~\ref{fig:plkSZgam3nuw0}.  It was found in  \citetalias{2018A&A...620A..78S} and \citetalias{2019A&A...631A..96I} that the neutrino mass alone has no effect on fixing the discrepancy; however, when considered along with $\gamma$, it was found that  $\gamma$ is then allowed to explore values that reduced the discrepancy, though remaining far from fixing it. Here when we follow the same scenario and allow massive neutrinos, but also a free $w$ along with $\gamma$, as expected we still see, as in  the case of a free $w$ and $\gamma$ with massless neutrinos, that the tension is clearly alleviated (brown lines). However, here again we lose this gain when we add BAO probes (orange lines), with the latter able to constrain both $w$ and the neutrino mass, leaving $\gamma$ alone unable to fix the discrepancy. Nevertheless, we note an improvement with respect to the case of a free $w$ and $\gamma$ with massless neutrinos, in the sense that the tension is reduced to the 2$\sigma$ level for a neutrino mass of $\sim 0.4$eV.

\subsection{Dynamical dark energy EoS vs a redshift-dependent growth index}

Here we focus, for generality and to allow more freedom for $w$ and $\gamma$, on investigating two additional cases. In  the first  the free dark energy EoS parameter evolves with time with two parameters $w_0$ and $w_a$ following the CPL parametrisation (Eq.~\ref{eq:CPL}) along with a constant $\gamma$, while in the second we instead allowed a redshift dependent growth index $\gamma$ following Eq.~\ref{eq:gamMar} along with  a free $w$ constant in time.
For the two schemes we use the parametrisation following Eq.~\ref{eq:sigmodgam} and as priors, $w_0\in [-3.0,-0.33]$ and $w_a\in [-4.0,3.0]$ while $\gamma_0\in [0.0,2.0]$ and $\gamma_1\in [-3.0,3.0]$. 

\begin{figure}[ht]
\centering
\includegraphics[width=\columnwidth]{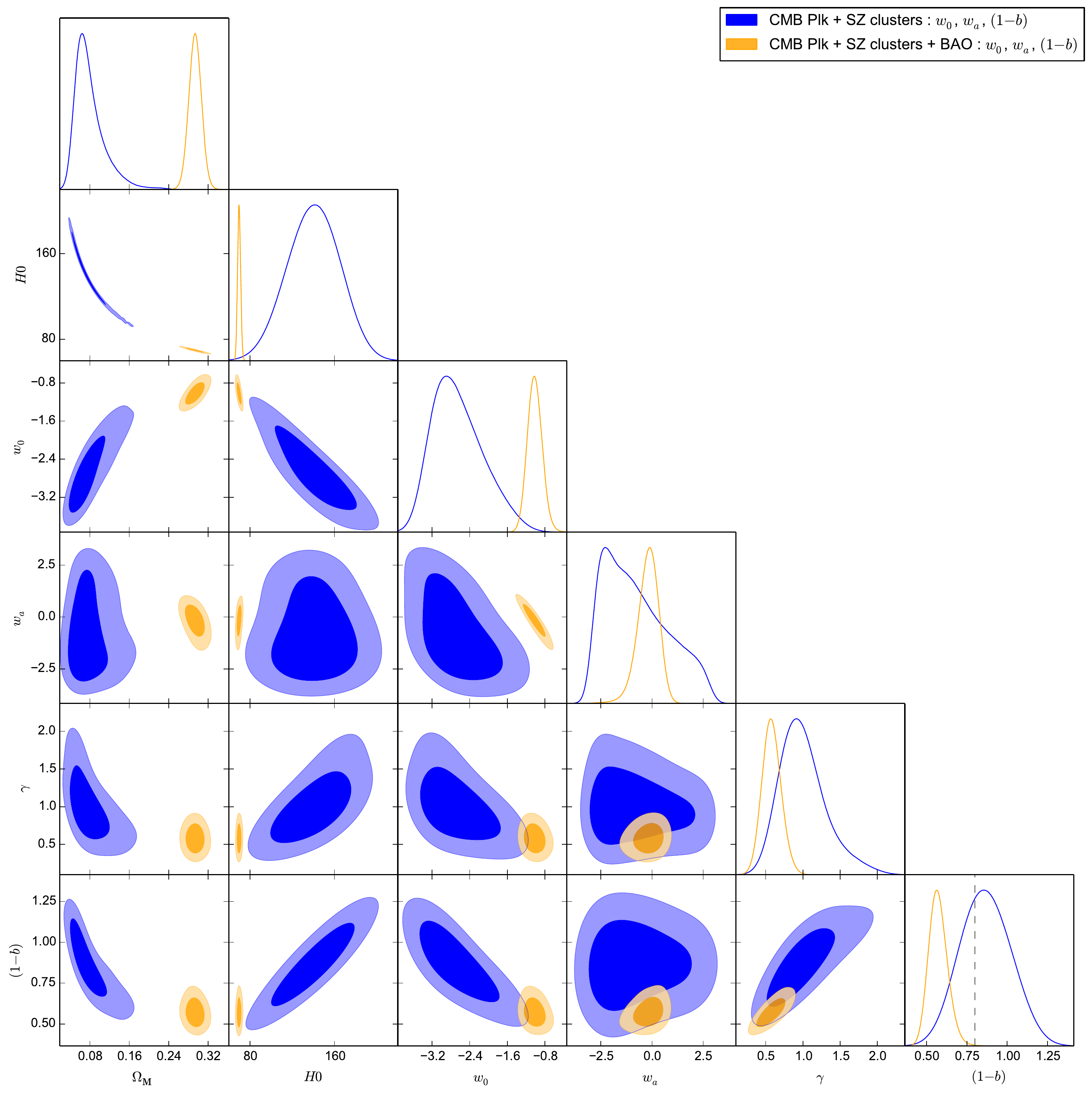}
\caption{Two-dimensional confidence contours at 68\%\ and 95\%  for the parameters $\Omega_{\rm m}$, $H_0$, $w_0$, and $w_a$ of a CPL parametrisation of dark energy EoS, $\gamma$, and $(1-b)$ derived from   Planck 2018 CMB datasets combined with the Planck 2015 SZ detected galaxy cluster sample (blue contours), then adding BAO probe datasets (orange contours). The vertical black dashed line gives  the Planck calibration value $\left (1-b \sim 0.8 \right )$.}
\label{fig:plkszbaogamaw0wa}
\end{figure}

In Fig.~\ref{fig:plkszbaogamaw0wa} we see that promoting the constant EoS parameter to one evolving with redshift with another degree of freedom $w_a$ has an impact on the discrepancy that is similar to the case when we allowed massive neutrinos along with a constant $w$ and $\gamma$, since the tension is largely alleviated (blue lines), with $1-b \sim 0.8$ falling at the maximum of likelihood. However, the addition of BAO tightens the confidence contours (orange lines) ruling out the Planck calibration, again by almost 3$\sigma$. Moreover, as observed before, even without adding BAO, solving the discrepancy would still need the matter density parameter as well as the Hubble constant to explore unrealistic values (i.e. excluded by almost all the cosmological probes) with $\Omega_{\rm m}$ heading below $0.1$ and $H_0$ exploring values higher than $100.0$ km/s/Mpc. 

\begin{figure}[ht]
\centering
\includegraphics[width=\columnwidth]{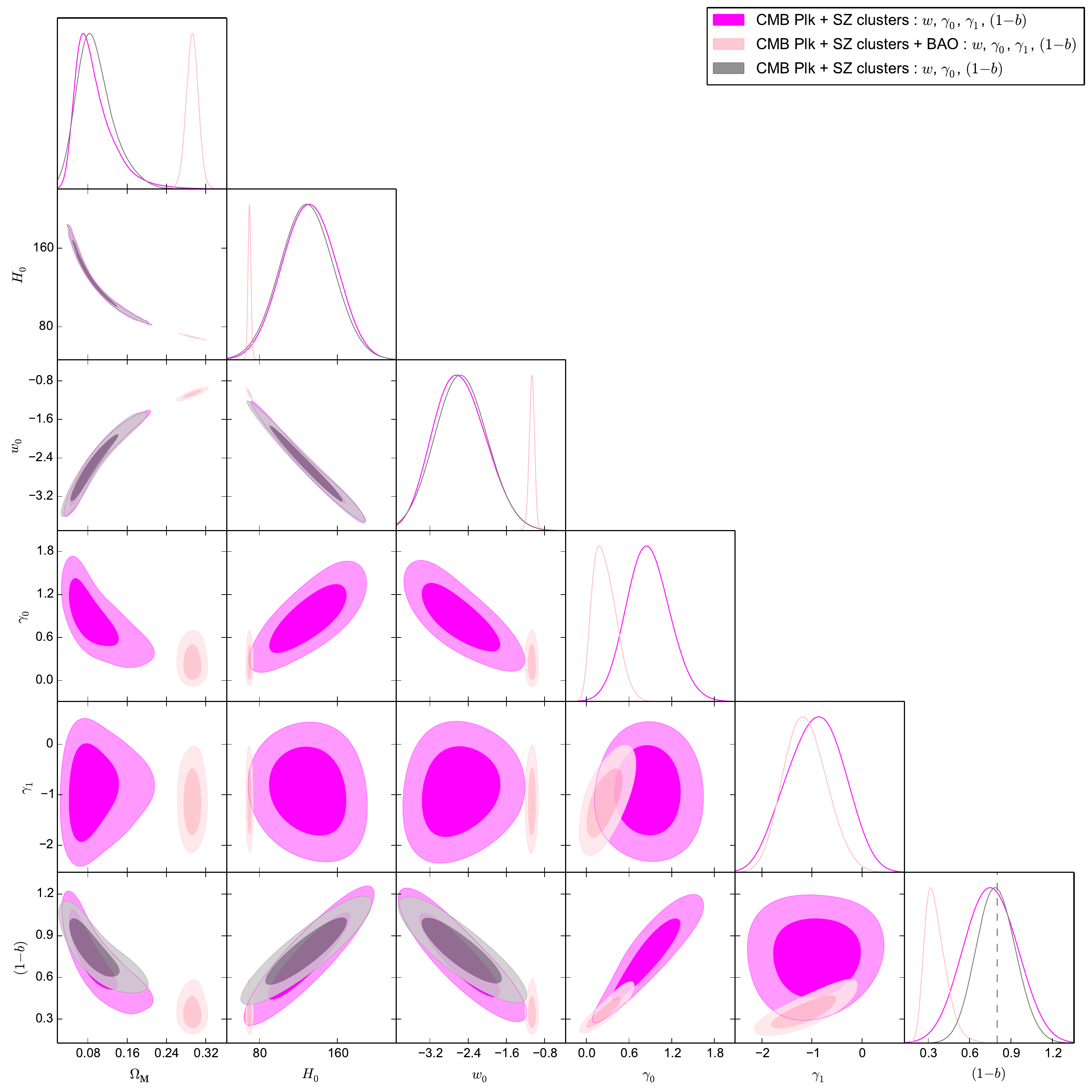}
\caption{Two-dimensional confidence contours at 68\%\ and 95\%  for the parameters $\Omega_{\rm m}$, $H_0$, $\gamma_0$,  $\gamma_1$, and  $w_0$ of a dynamical parametrisation of the growth index and $w_0$ and $(1-b)$, derived from Planck 2018 CMB datasets combined with Planck 2015 SZ detected galaxy cluster sample, then combined with BAO probe datasets. The vertical black dashed line gives the Planck calibration value $\left (1-b \sim 0.8 \right )$.}
\label{fig:plkSZbalphagam0gam1w0}
\end{figure}

The situation does not change when we consider instead a dynamical $\gamma$ as in Fig.~\ref{fig:plkSZbalphagam0gam1w0}. The additional $\gamma_1$ degree of freedom only results in  slightly larger confidence contours (magenta lines) with respect to the case of a constant growth index with a free $w$ (grey lines), while the  $\Omega_{\rm m}$ and $H_0$ constraints are outside the common values. Moreover, when we combine with BAO, although it  is not able to further constrain the growth index,  we observe  that an evolving growth index drives $(1-b)$ towards lower values (pink lines) that would rather increase the discrepancy on $\sigma_8$.

\section{Further tests of evidence}\label{sec:furthertest}

In this section we try to further consolidate the previous conclusions by adopting a model comparison approach, i.e. we consider that we are comparing two models, the one with the free calibration parameter $(1-b)$ with the three main extensions to $\Lambda$CDM ($w$, $w$ + $\gamma$, and $w$ + $\gamma$ + $\sum m_{\nu}$) against the one with $(1-b) \sim 0.8$, the calibration adopted by Planck, allowed to vary but with a very narrow Gaussian error $\pm 0.01$ following \citep{2014A&A...571A..29P}, also considering the previous extensions. We then compare the Bayesian MCMC outcomes for the two models by the classical 1D and 2D likelihood confidence contour separation in $\sigma$ and by using another method through the Bayesian evidence index (see Sect.~\ref{sec:impL} for a description and the method of evaluation).

\begin{figure}[ht]
\centering
\includegraphics[width=0.55\columnwidth]{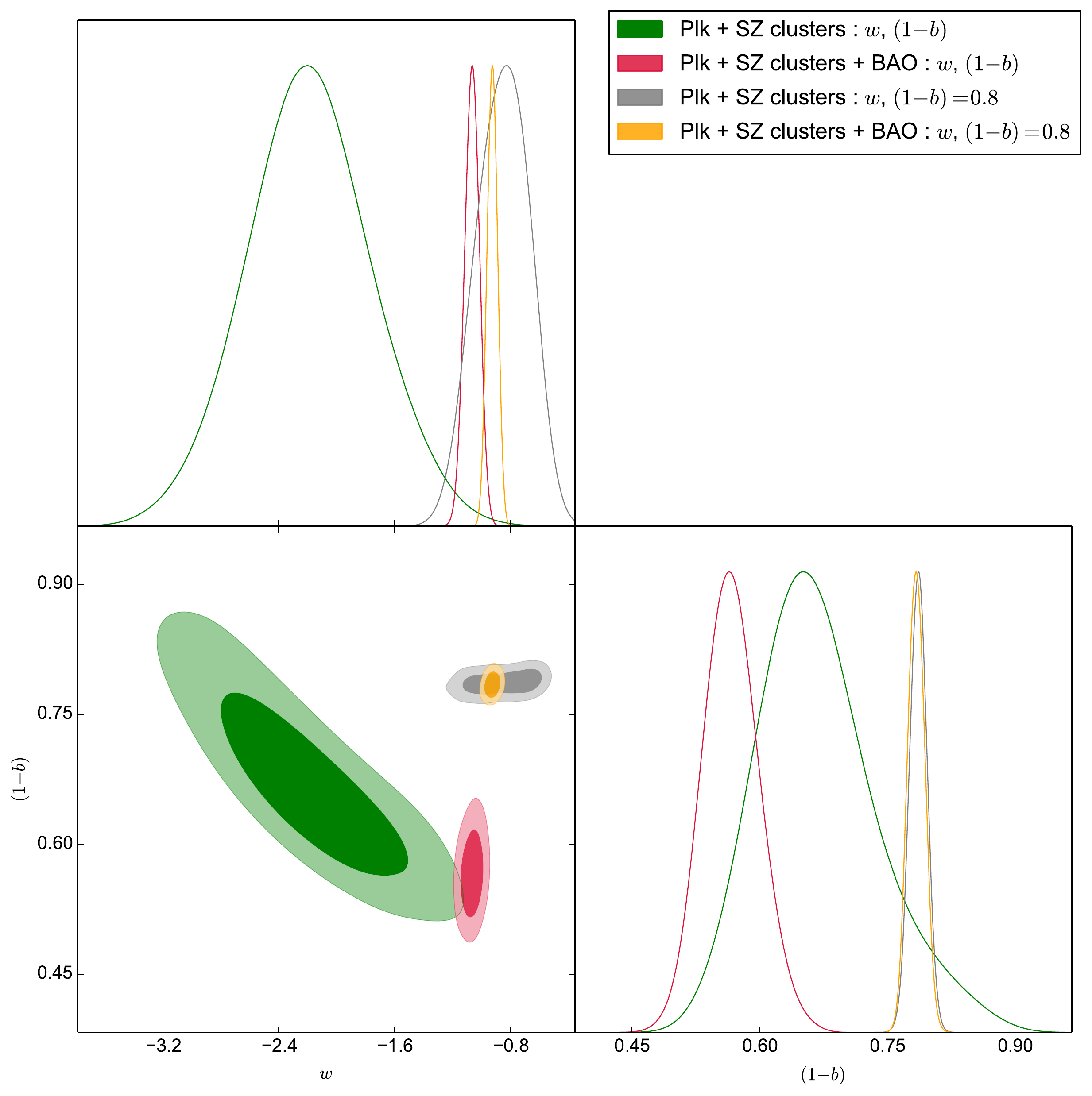}
\includegraphics[width=0.55\columnwidth]{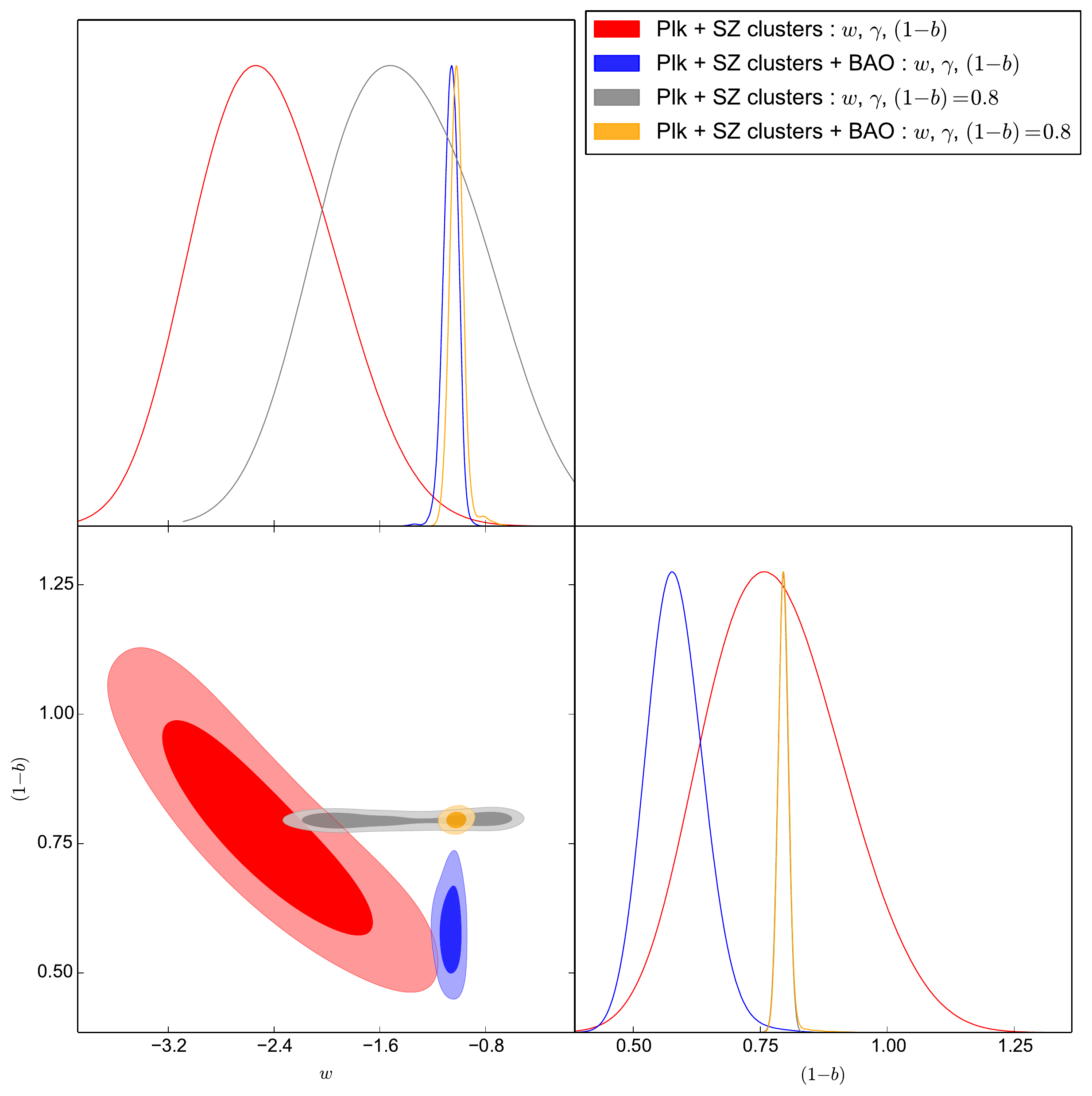}
\includegraphics[width=0.55\columnwidth]{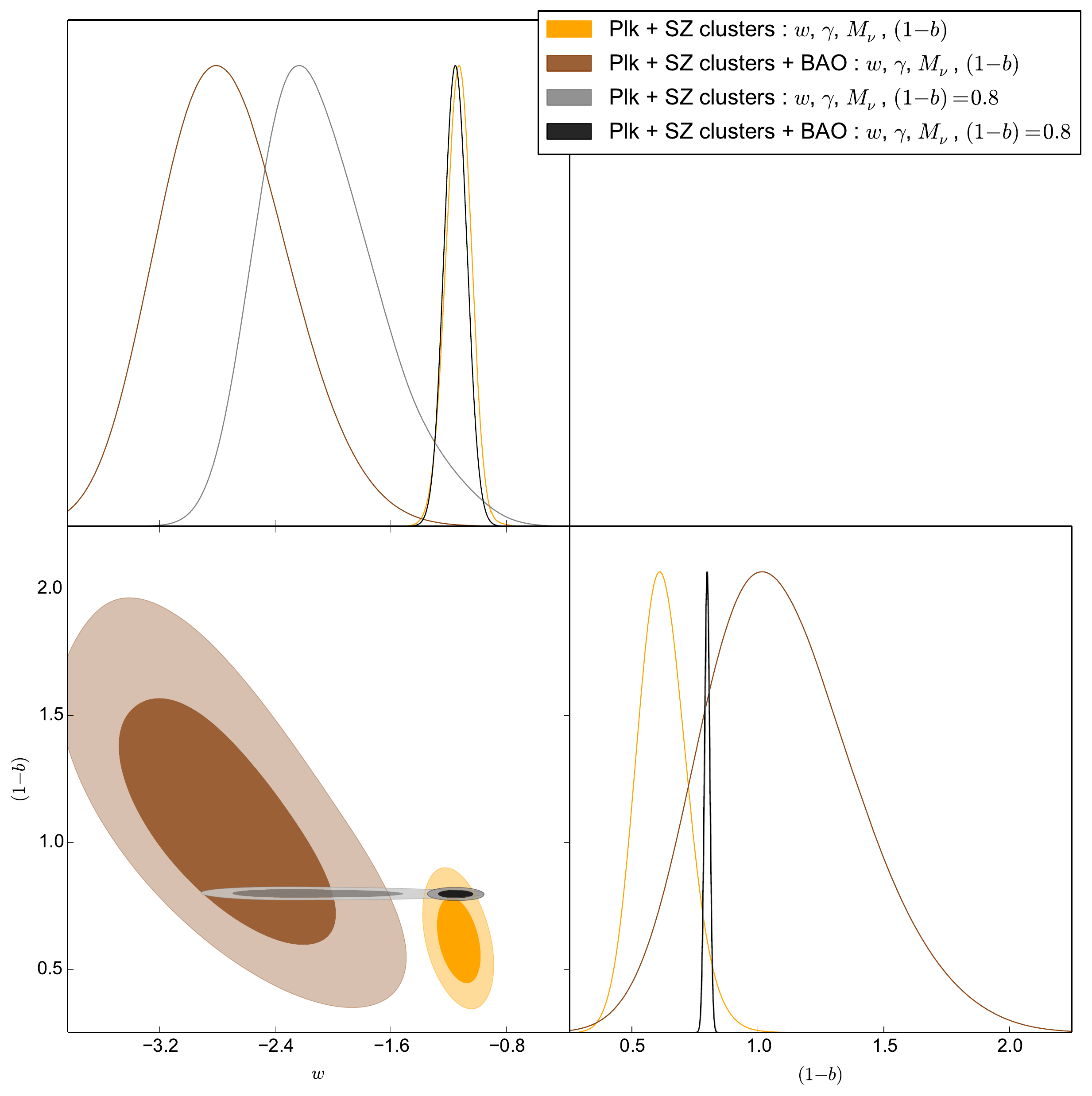}
\caption{Confidence contours (68 and 95\%) and posterior distributions for the dark energy EoS parameter $w$ and mass calibration parameter $(1-b)$ when combining CMB (with or without BAO) and SZ cluster data for three different cases of allowing extensions to $\Lambda$CDM (detailed in the legends), in comparison with confidence contours from the effect of a strong prior on $(1-b)$ centred around the Planck calibration value of 0.8.}
\label{fig:plkSZbwgamnub08}
\end{figure}

We thus show in Fig.~\ref{fig:plkSZbwgamnub08} the confidence contours of the $(1-b)$-$w$ parameters for the three $\Lambda$CDM extensions for the two cases: a free calibration parameter and $1-b\sim 0.8$ with a very restrictive Gaussian prior, and that using the combination of clusters and CMB datasets along with the same scheme but adding the  BAO data. We clearly see in the first panel that the Planck calibration is not reached even with a free $w$, while imposing $1-b \sim 0.8$ restricts $w$ closer to -1. Adding BAO further consolidates  the finding by shrinking the green contours to the red ones in  Fig.~\ref{fig:plkSZbwgamnub08}. Allowing in addition a free $\gamma$ in the second panel improves the situation and reduces to 1~$\sigma$ the discrepancy when comparing the red and grey contours, but the addition of BAO increases it again to $\sim$3~$\sigma$. Instead,  further allowing neutrinos at the expense of more degrees of freedom, moves more towards alleviating the tension, even if we assume a very good accuracy on $(1-b)$, with the grey contours overlapping   the brown ones. However, still combining with BAO restores it back to $\sim$2~$\sigma$. 

The values of the the Bayes factor in Table \ref{tab:plkSZbwgamnub08} reflects the previous observations. Thus, we see in the first (respectively second) column, corresponding to the Bayes factor computed for the three extensions for the cases without (respectively with) BAO, that $\ln B_{ij}$ is always > 1, which we consider   strong evidence that the model with $1-b \sim 0.8$ is not preferred.  We see that  $\ln B_{ij}$  decreases when considering further extensions, which  is a sign that the addition of these extensions pushes towards the model with the Planck calibration,  however without preferring it. We also observe that the combination with BAO in $\Lambda$CDM increases $\ln B_{ij}$ as expected and depicted in Fig.~\ref{fig:plkSZbwgamnub08} with the exception of the case with only free $w$. The reason for this might be that BAO tightly restricts $w$ to the same range for the two models, which is reflected in the calculation of the Bayes factor that includes global contribution from the parameters involved.  These parameters are not compensated for by the difference in the 1D likelihood of $(1-b)$, unlike the other two extensions where the higher overlapping of the 1D confidence contours of the $(1-b)$ parameters cover the gain from restricting $w$. 

\begin{table}\label{tab:plkSZbwgamnub08}
\begin{center}
\setlength{\tabcolsep}{5pt}
\renewcommand{\arraystretch}{1.55}
\caption{Bayes factor ($\ln B_{ij}$) values for the $\Lambda$CDM extensions and mass observable parameters. }
\begin{tabular}{ccc}
\hline
\hline
Parameters \textbackslash \, Data & CMB + Clust  & CMB + Clust + BAO \\ \hline
$w$            & $20.64$ & $19.79$ \\ \hline
$w$ + $\gamma$ & $4.42$ & $5.89$ \\ \hline
$w$ + $\gamma$ + $\sum m_{\nu}$ & $1.81$ & $3.77$ \\ \hline
\end{tabular}
\end{center}
\end{table}


\section{Discussion and conclusion}
\label{sec:concL}

In the present paper we examined whether a dark energy EoS parameter $w$ is able to fix the discrepancy found in the value of $\sigma_8$ determined from local probes like cluster counts with respect to that obtained at deep redshifts from CMB $C_{\ell}$. Our strategy was to examine the constraints that CMB and cluster abundance data yield, without further additional assumption on clusters (i.e. leaving the calibration parameter of the mass observable relation free to vary). In most of the cases the addition of massive neutrinos was also considered. We found first, independently from allowing or not massive neutrinos, that a dark energy EoS $w$, constant or evolving with time, when left free, can fix the discrepancy only for $w$ well below $-1$. However, these values are excluded by additional constraints from BAO data.\\
Next we introduced, along with a free $w$, the $\gamma$ parametrisation to model the effect of a modification of gravity on the growth rate. Though the discrepancy is further reduced with the $(1-b)$ maximum likelihood value becoming centred around 0.8; however, here again, combining it with geometrical distance probes like BAO reverts the $(1-b)$ preferred value back to the  $\Lambda$CDM value of 0.6. The same is observed after introducing, along with $w$ and $\gamma$, further extensions to $\Lambda$CDM like massive neutrinos, where we found that a greater ease to the tension is found for higher neutrino masses only when they are not constrained by the combination with BAO datasets.\\
Finally, we compared the effect of allowing an evolving $w$ in time through CPL parametrisation in addition to constant growth index against the case where the index is made dynamical through a second parameter $\gamma_1$ in addition to a constant $w$. In the first case, we reached the same conclusions as with a constant $w$ and $\gamma$, where the discrepancy was alleviated only if we do not further constrain $w$ by BAO, while in the second case we find that  an evolving growth index does not bring any improvement to the situation since it drives $(1-b)$ towards lower values that instead increase the discrepancy on $\sigma_8$.\\
To further consolidate  our findings, we  compared the models with the extensions and a free calibration parameter against those with the same extensions but with a calibration considered to be constrained to the per cent level  to the Planck calibration, and that by means of the common method of comparing the confidence contours from MCMC chains or through another method, based on comparing the values of the Bayes factors for each case. We found that the two methods are in agreement with our previous findings and that the values of the Bayes factor strongly supports that the proposed extensions are not preferred over a change in the value of the mass observable calibration as a way to fix the discrepancy.    

We conclude then that, none of the extensions to $\Lambda$CDM model we tested as possible solutions to solve the discrepancy is preferred over the hypothesis of an improper calibration of the mass observable scaling relation. However, if the Planck SZ calibration,  $(1-b)\sim0.8$, is consolidated in the future (e.g. to $\sim$ 1\% in the case we showed), the failure of the advocated extensions to $\Lambda$CDM to fix the discrepancy (i.e. massive neutrinos, dark energy, and a modified gravity through $\gamma$) would call for new physics beyond the standard cosmological model common modifications. 

This is especially true after finding, within the probes used in this work, that the constraints on the influential parameters need to drastically shift if we want to ease the tension :  the matter density or dark energy EoS parameter should reach very low values, around $\sim 0.1$ for $\Omega_m$ and $<-3$ for $w$, while the Hubble constant and the growth index prefer very high values, $>100$ for $H_0$ and $\sim 1.0$ for $\gamma$. This puts the discrepancy at a high level in terms of the difficulty to solve, without spoiling the remarkable concordance of the $\Lambda$CDM model (see e.g. \cite{Benevento:2022cql} for a relative success in doing so with a modified gravity model). Within the current available measurements, in addition to the numerous studies reviewed in the introduction of this work (see Sect.~\ref{sec:IntrO}) there have been some recent attempts to investigate and improve the bounds on the mass observable calibration, such as using machine learning methods to reduce the SZ flux-mass scatter   \citep{Wadekar:2022cyw}, but only future generation surveys, such as Euclid  \cite{EUCLID:2011zbd}, eRosita \cite{Borm:2014zna}, DESI \cite{DESI:2016fyo}, Wfirst-Roman \cite{Spergel:2015sza}, LSST-Rubin \cite{Abell:2009aa}, SKA \cite{SKA:2018ckk}, and NIKA2 SZ Large Program (LPSZ) \cite{Keruzore:2021jdc}, will allow us to reach  a high reduction of the error on the calibration of the mass observable that will confirm or not this discrepancy to a very high degree of evidence (see also \citealt{Pratt:2019cnf} for a review on the latest techniques and future prospects).

\bibliographystyle{aa}
\bibliography{Bibliography}
\begin{appendix}

\section{Hubble constant prior range and its impact on the posteriors}

\begin{figure}[h!]
\centering
\includegraphics[width=\columnwidth]{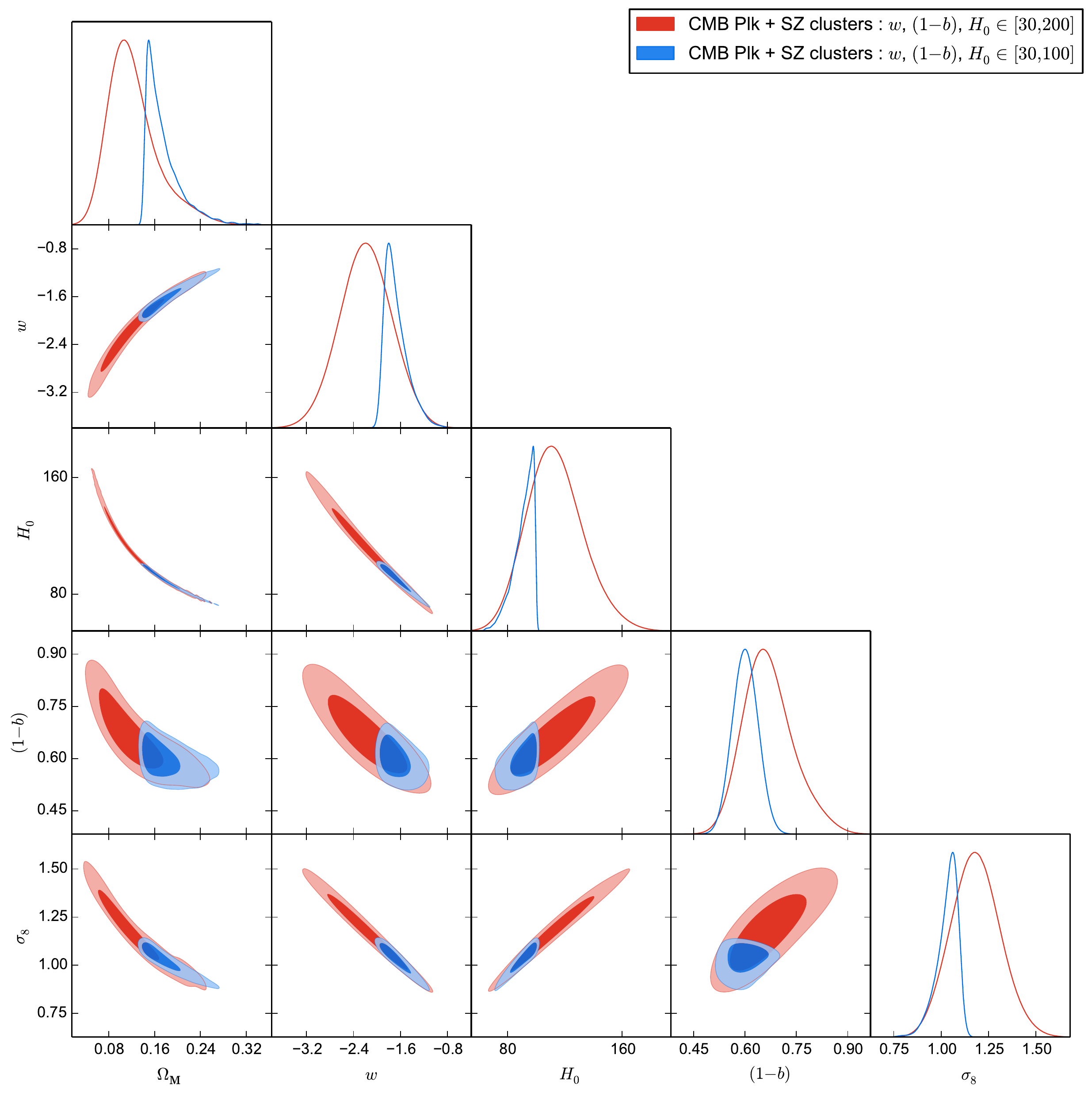}
\caption{Two-dimensional confidence contours at 68\%\ and 95\%  for the parameters $\Omega_{\rm m}$, $\sigma_8$, $(1-b)$, and $w$ derived from Planck 2018 CMB datasets combined with the  Planck 2015 SZ detected galaxy cluster sample for $H_0 \in [30,100]$ (blue lines) in comparison with $H_0 \in [30,200]$ (red lines).}
\label{fig:plkjlaSZATM3nuwo_2}
\end{figure}

\begin{figure}[h!]
\centering
\includegraphics[width=\columnwidth]{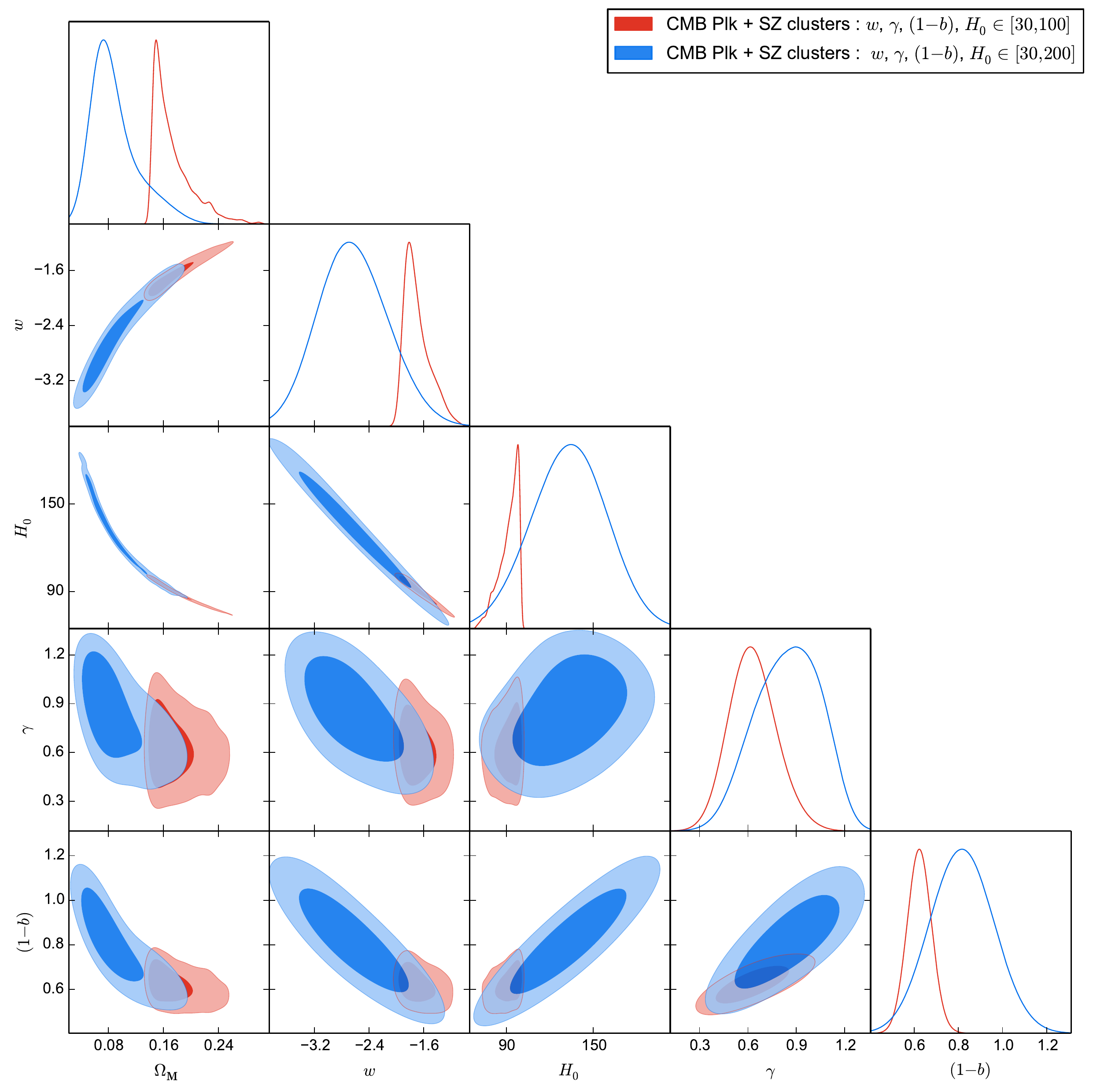}
\caption{Confidence contours at 68\%\ and 95\%  for the parameters $(1-b)$ , $\sigma_8$, $\Omega_{\rm m}$, $w$, and $\gamma$ derived from Planck 2018 CMB datasets combined with the  Planck 2015 SZ detected galaxy cluster sample for $H_0 \in [30,100]$ (red lines) in comparison with $H_0 \in [30,200]$ (blue lines).}
\label{fig:plkSZgam3nuw0_2}
\end{figure}

So far, when using only CMB datasets and cluster counts, we allowed large priors for the Hubble constant enabling it to potentially reach $H_0 \sim$ 200.0, well above the current constraints from other probes like BAO for example. The motivation was to permit $w$ to explore values well below $-1$ that may reconcile CMB and cluster counts constraints on $\sigma_8$. We observed in Fig.~\ref{fig:plkjlaSZATM3nuwo} that the discrepancy on $\sigma_8$, which directly translates into that on $(1-b)$, was alleviated for $w\ll$ -1 where the  $(1-b)$ likelihood reached $\sim$ 0.8, the Planck SZ sample calibration, to less than 2$\sigma$ and  for values of $H_0 \sim$ 160.0. The same was also observed in other cases, in particular in Fig.~\ref{fig:plkSZgam3nuw0} where we also allowed  the growth index $\gamma$ to vary. Here we repeated, in Fig.~\ref{fig:plkjlaSZATM3nuwo_2} and Fig.~\ref{fig:plkSZgam3nuw0_2}, the  previous analysis, but  limiting the prior range of the Hubble constant to $\in [30,100]$, in line with the current constraints on $H_0$. We observe then three main differences between the two posteriors for the two considered $H_0$ priors:
\begin{itemize}
\item The discrepancy is not alleviated  in the free $w$ only case with (1-b) maximum likelihood centred far from $1-b\sim 0.8$ with the latter falling at a level of more than 3$\sigma$.  The same is seen in the free $w$ and $\gamma$ cases where the Planck calibration still falls at the 3$\sigma$ level. This is especially seen in the $w$ plus $\gamma$ case where the \textit{\emph{cut}} in the  $H_0$ prior does not trivially translate in the posterior contours with also a change in the likelihood maximum value.
\item The preferred values of $w$ in the $H_0$ tight prior case are now close to -1, while they were far below that value for the  $H_0$ large prior case with the $w$ corresponding to $\Lambda$CDM falling at the $3\sigma$ level.
\item The correlation between $w$ and $(1-b)$, found in the $H_0$ large prior case as a sign of its ability to fix the discrepancy, has vanished in the  tight prior case, while that of the modified gravity related parameter $\gamma$ with $(1-b)$ is still strong, indicating its ability, unlike $w$, to solve the discrepancy. This can also be seen in Table~\ref{SZtable} where the correlation coefficient described in Sect.~\ref{sec:impL} is calculated for different cases with $h \in [30,100]$. We then see that the index for $(1-b)$ - $\gamma$ reaches high values, above 0.5, contrary to $(1-b)$ - $w_0$, which remains at very low values, well below the median for the correlation index. \\
\end{itemize}

\begin{table}[t]
\caption{Evaluations of the Pearson correlation coefficient for the pair $(1-b)$ - $\gamma$ vs $(1-b)$ - $\omega_0$, the  inferred parameters from MCMC using different combinations of the CMB datasets and the SZ detected cluster sample.}\label{SZtable}
\resizebox{\columnwidth}{!}{
\begin{tabular}{c|c|c|}
\cline{2-3} &  $(1-b)$ - $\gamma$ & $(1-b)$ - $\omega_0$  \\

\cline{1-3}

\multicolumn{1}{|l|}{} & $$ & $$ \\

\multicolumn{1}{|l|}{SZ+CMB} & $0.711$ & $0.295$ \\[0.1cm]

\multicolumn{1}{|l|}{SZ+CMB+$M_{\rm \nu}$} & $0.736$ & $-0.017$ \\[0.1cm]

\multicolumn{1}{|l|}{SZ+CMB+BAO} & $0.750$ & $-0.010$   \\[0.1cm]
 
\multicolumn{1}{|l|}{SZ+CMB+$w_0-w_a$} & $0.749$ & $-0.024$   \\[0.1cm]
 
 \multicolumn{1}{|l|}{SZ+CMB+$\gamma_0-\gamma_1$} & $0.702$ & $0.241$   \\
 \hline

\end{tabular}
}
\end{table}

\section{Effective relativistic number of neutrinos in combination with $w$}\label{effNeff}

\begin{figure}[h!]
\includegraphics[width=\columnwidth]{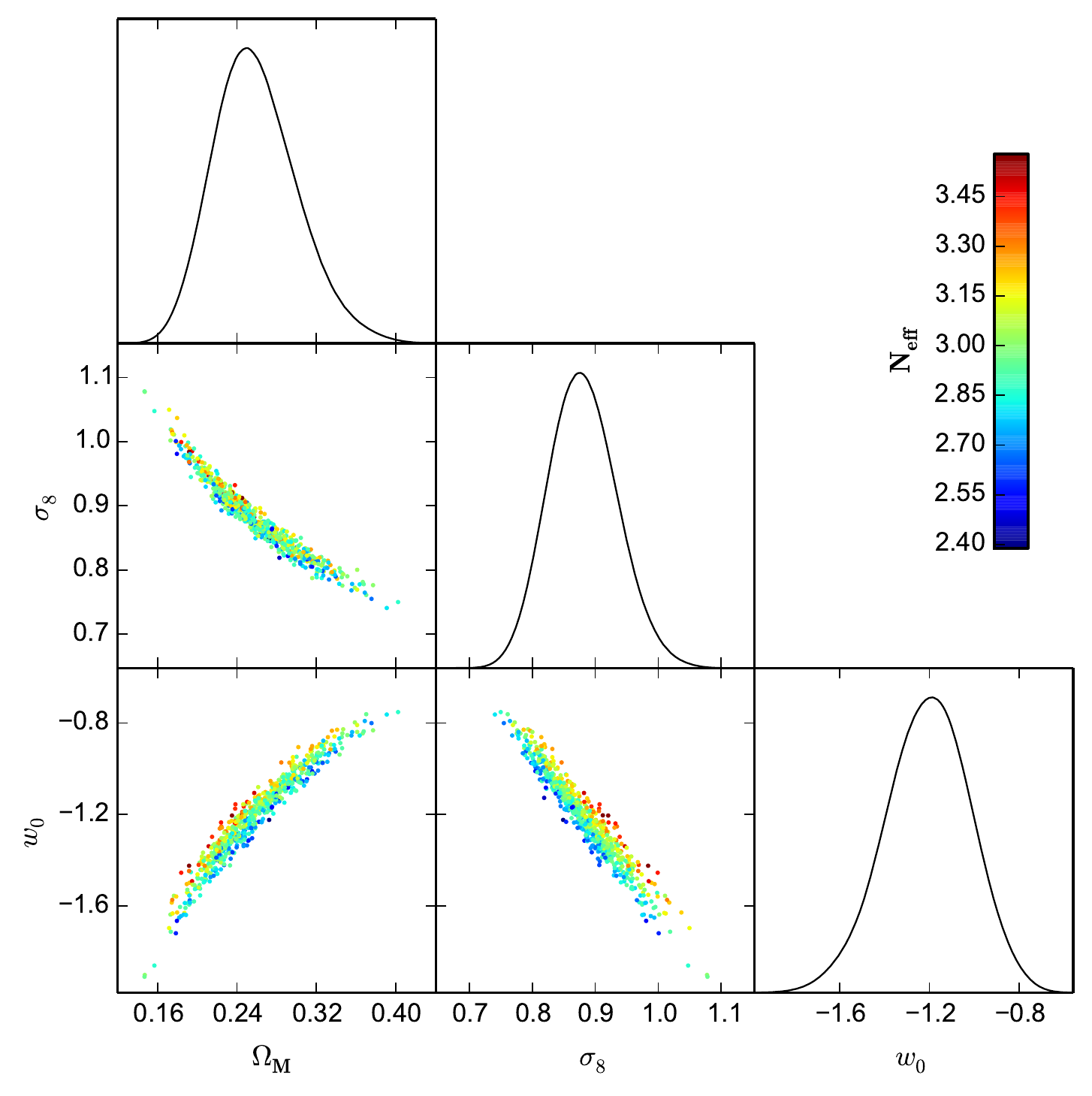}
\caption{Two-dimensional confidence contours at 68\%\ and 95\%  for the parameters $\Omega_{\rm m}$, $\sigma_8$, $w_0$, free constant dark energy, and the effective number of neurinos $N_{\rm eff}$ in color distribution representation derived from MCMC analysis combining CMB datasets and a local X-ray cluster sample.} 
\label{fig:BAXGATMNeffw0cut3d}
\end{figure}

In the main text we showed that allowing massive neutrinos has no substantial effect in helping $w$ to fix the discrepancy. This was while keeping the number of neutrino species  $N_{\rm eff}\sim 3.046$, the value favoured by earth experiments. Here we leave $N_{\rm eff}$ free in addition to $w$ while fixing the mass calibration parameter to the Planck calibration. We see in Fig.~\ref{fig:BAXGATMNeffw0cut3d} that the  $N_{\rm eff}$ confidence contours are in agreement with the fiducial value $3.046$ as well as $w$ with -1,   mainly because the variation in $N_{\rm eff}$ induce changes to the correlation $w$ - $\Omega_{\rm m}$ in opposite directions to that between $w$ and $\sigma_8$, as indicated from the colour scheme  we used to represent the variation in $N_{\rm eff}$. This results in keeping the $\sigma_8-\Omega_{\rm m}$ contours unchanged and in agreement with fiducial $N_{\rm eff}$, clearly shows that varying the  $N_{\rm eff}$ parameter does not help $w$ in reconciling cluster counts and  CMB $C_{\ell}$ constraints on $\sigma_8$.\\
\end{appendix}
   
\begin{acknowledgements}
We thank the anonymous referees for their valuable and constructive comments and suggestions on our manuscript. Ziad Sakr was supported by a grant
of excellence from the Agence des Universités Francophones (AUF). Ziad Sakr acknowledges support from the IRAP Toulouse and IN2P3 Lyon computing centers. Stéphane Ilic thanks the Centre national d'études spatiales which supports his postdoctoral research contract. 
\end{acknowledgements}

%
%

\end{document}